\documentclass[prl,twocolumn,letterpaper,byrevtex,floatfix,%
preprintnumbers,superscriptaddress,amsfonts,amssymb,amsmath]{revtex4}

\usepackage{mathptmx}
\usepackage{url}
\usepackage{hyperref}
\usepackage{graphicx}

\newcommand*{\CThree}{\ensuremath{\mathbb{C}^3}}

\newcommand*{\RThree}{\ensuremath{\mathbb{R}^3}}
\newcommand*{\pddt}[1]{\frac{\partial{#1}}{\partial t}}
\newcommand*{\I}{\mathrm{i}}
\newcommand*{\Sub}[1]{_{\mathrm{#1}}}
\newcommand*{\Sup}[1]{^{\mathrm{\,#1}}}
\newcommand*{\Eg}{\emph{E.g.}}

\newcommand*{\ie}{\emph{i.e.}}

\renewcommand*{\Im}[1]{\mathinner{\mathrm{Im}\left\{{#1}\right\}}}
\renewcommand*{\Re}[1]{\mathinner{\mathrm{Re}\left\{{#1}\right\}}}

\newcommand*{\Mag}[1]{\mathinner{\left|{#1}\right|}}
\newcommand*{\Tensor}[1]{\ensuremath{\boldsymbol{\mathsf{#1}}}}
\renewcommand*{\vec}[1]{\ensuremath{\mathbf{#1}}}
\newcommand*{\cross}{\boldsymbol{\times}}
\newcommand*{\del}{\boldsymbol{\nabla}}
\renewcommand*{\div}{\del\cdot}
\newcommand*{\curl}{\del\cross}

\newcommand*{\B}{\vec{B}}
\newcommand*{\E}{\vec{E}}
\newcommand*{\f}{\vec{f}}
\newcommand*{\fLorentz}{\f\Sup{Lorentz}}
\newcommand*{\fRS}{\f\Sup{RS}}
\newcommand*{\fspin}{\f\Sup{spin}}

\newcommand*{\epsz}{\varepsilon_0}
\newcommand*{\G}{\vec{G}}
\renewcommand*{\j}{\vec{j}}
\newcommand*{\J}{\vec{J}}

\newcommand*{\K}{\vec{K}}

\newcommand*{\N}{\vec{N}}
\newcommand*{\p}{\vec{p}}
\newcommand*{\pfield}{\p}

\newcommand*{\s}{\vec{s}}

\newcommand*{\V}{\vec{V}}
\newcommand*{\x}{\vec{r}}
\newcommand*{\0}{\vec{0}}

\begin{document}
\preprint{}

\title{%
Conservation laws in generalized Riemann-Silberstein electrodynamics%
}

\author{Jan E.~S.~Bergman}
\email{jb@irfu.se}
\affiliation{%
Swedish Institute of Space Physics,
P.\,O.\,Box 537,
SE-751\,21 Uppsala,
Sweden}%

\author{Siavoush M. Mohammadi}
\affiliation{%
Swedish Institute of Space Physics,
P.\,O.\,Box 537,
SE-751\,21 Uppsala,
Sweden}%

\author{Lars~K.~S.~Daldorff}
\affiliation{%
Department of Physics and Astronomy,
Uppsala University,
P.\,O.\,Box 516,
SE-751\,20 Uppsala,
Sweden}%

\author{Bo Thid{\'e}}
\altaffiliation[Also at ]{LOIS Space Centre, V\"axj\"o University,
 SE-351\,95 V\"axj\"o, Sweden}
\affiliation{%
Swedish Institute of Space Physics,
P.\,O.\,Box 537,
SE-751\,21 Uppsala,
Sweden}%

\author{Tobia D.~Carozzi}
\affiliation{%
Department of Physics and Astronomy,
University of Glasgow,
Glasgow G12 8QQ Scotland,
United Kingdom}%

\author{Roger L. Karlsson}
\affiliation{%
Space Research Institute,
Austrian Academy of Sciences,
Schmiedlstra{\ss}e 6,
A-8042 Graz,
Austria}%
\affiliation{%
Department of Physics and Astronomy,
Uppsala University,
P.\,O.\,Box 516,
SE-751\,20 Uppsala,
Sweden}%

\author{Marcus Eriksson}
\affiliation{%
Swedish Institute of Space Physics,
P.\,O.\,Box 537,
SE-751\,21 Uppsala,
Sweden}%
\affiliation{%
Department of Mathematics,
Uppsala University,
P.\,O.\,Box 480,
SE-751\,06 Uppsala,
Sweden}%

\begin{abstract}

Starting from positive and negative helicity Maxwell equations expressed in
Riemann-Silberstein vectors, we derive the ten usual and ten
additional Poincar{\'e} invariants, the latter being related to the
electromagnetic spin, \ie, the intrinsic rotation, or state of
polarization, of the electromagnetic fields.  Some of these invariants
have apparently not been discussed in the literature before.

\end{abstract}


\maketitle

The electromagnetic angular momentum and its relation to photon spin
have been much debated
\cite{Humblet:Physica:1943,Jauch&Rohrlich:Inbook:1980,vanEnk&Nienhuis:EPJ:1994,vanEnk&Nienhuis:JMO:1994}.
To help resolve this issue, we have derived conservation laws for
Maxwell electrodynamics by using a generalized Riemann-Silberstein (RS)
vector formalism \cite{Silberstein:AP:1907a,Silberstein:AP:1907b} that
incorporates optical coherence theory
\cite{Mandel&Wolf:Book:1995,Cohen:Book:1995}.  In terms of RS vectors
$\G_\pm = \E \pm \I c\B$, where $\E,\B\in\CThree$ are typically
represented in a complex helical base, the Maxwell equations are
\begin{align}
\label{eq:divG}
\div\G_\pm &=\frac{\rho}{\epsz}\,,\\
\label{eq:curlG}
\curl\G_\pm &= \pm \I\left(\frac{1}{c}\pddt{\G_\pm}+Z_0\j\right)\,,
\end{align}
where $\rho,\j\in\CThree$ are the charge and current densities,
respectively, and $Z_0=\sqrt{\mu_0/\epsz}$ is the vacuum impedance.
Defining the electromagnetic energy and momentum densities as
$H_{\pm}=\epsz\G_\pm\cdot\G_\pm^\ast/2$ and
$\K_\pm=\mp\epsz\Im{\G_\pm\cross\G_\pm^\ast}/(2c)$, respectively, one
can easily derive the conservation laws
\begin{gather} 
\label{eq:GRSenergy} 
\pddt{H_{\pm}}+c^2\div\K_\pm+\Re{\j\cdot\G_\pm^\ast}= 0\,,\\
\label{eq:GRSmomentum}
\pddt{\K_\pm} +\div\tilde{\Tensor{T}}_\pm + \fRS = \0\,,
\end{gather}
where $\fRS_\pm=\Re{c\rho\G_{\pm}^\ast\pm
\I{}\j\cross\G_\pm^\ast}/c$ are electromechanical force densities,
coined Riemann-Silberstein force densities, and the stress tensors
are $\tilde{T}_\pm^{ij}=\delta^{ij}H_\pm-\epsz\Re{G_\pm^{i}G_\pm^{j\ast}}$.

Introducing the field energy density
$u=\epsz(\E\cdot\E^\ast+c^2\B\cdot\B^\ast)/2$;
the linear momentum density
$\pfield=\epsz\Re{\E\cross\B^\ast}$;
the Maxwell stress tensor
$\Tensor{T}=(\tilde{\Tensor{T}}_{+}+\tilde{\Tensor{T}}_{-})/2$;
the Lorentz force density
$\fLorentz=\Re{\rho\E^{\ast}+\j\cross\B^\ast}$;
the spin energy density
$w=\Im{\E\cdot\B^\ast}/Z_0$;
the spin momentum density
$\V=-\epsz\Im{\E\cross\E^\ast+c^2\B\cross\B^\ast}/(2c)$;
the spin force density
$\fspin =\Im{c\rho\B^\ast-\j\cross\E^\ast/c}$;
and the spin stress tensor 
$\Tensor{U}=(\tilde{\Tensor{T}}_+-\tilde{\Tensor{T}}_-)/2$,
summing and subtracting the two conservation laws in
(\ref{eq:GRSenergy}), and summing and subtracting the two conservation
laws in (\ref{eq:GRSmomentum}), we obtain the following conservation laws
\begin{gather}
\label{eq:Poynting}
\pddt{u}+c^2\div\pfield + \Re{\j\cdot\E^\ast} = 0\,, \\
\label{eq:lincon}
\pddt{\pfield}+\div{\Tensor{T}} + \fLorentz = \0\,, \\
\label{eq:spin-energy}
\pddt{w} + \div\V +\Im{\j\cdot\B^\ast} = 0\,, \\
\label{eq:spincon}
\pddt{\V}+\div{\Tensor{U}} + \fspin= \0\,,
\end{gather}
respectively
\cite{Bogoliubov&Shirkov:Book:1959,Kujawski:NC:1966,Carozzi&al:PRE:2000}.
The conservation laws (\ref{eq:spin-energy}) and (\ref{eq:spincon})
involving the spin momentum density pseudovector $\V$, which is a
three-dimensional generalization of the standard 2D Stokes parameter $V$
\cite{Brosseau:Book:1998,Carozzi&al:PRE:2000}, seem to have gone largely
unnoticed until now.  Note that for linearly polarized fields,
$\E,\B\in\RThree$ and $\V=\0$.  For circularly polarized fields
$\E=\pm\I{}c\B$ and ${\pfield=\pm\V}$.  If the field is monochromatic,
then the spin angular momentum (SAM) density is $\s=c\V/\omega$ and the
electromagnetic spin torque density is
${\boldsymbol\tau}\Sup{spin}=c\fspin/\omega$
\cite{Poynting:PRSL:1909,Beth:PR:1936,Carrara:NC:1949}.

Using the conservation laws Eqs.~(\ref{eq:GRSenergy}) and
(\ref{eq:GRSmomentum}), the energy and momentum can be written as
$H_\pm=u\pm v$, $\K_\pm=\pfield\pm\V$ and $\fRS_\pm=\fLorentz\pm\fspin$.
This elucidates the difference in handedness between $\G_+$ and $\G_-$,
which are linearly independent under Lie transformations
\cite{Nikitin&al:Incoll:1985}.  We interpret $\G_\pm$ as wave fields
of positive and negative helicity $\chi=\V\cdot\pfield/\Mag{\pfield}^2$,
manifesting the well-known fact that left and right handed polarized
fields represent two different helicities \cite{Birula:PO36:1996};
$\chi-\pm1$ for $\E=\pm{}\mathrm{i}c\B$.

The interpretation of the RS vector as a photon wave function was
suggested by many authors
\cite{Birula:APP:1994,Sipe:PRA:1995,Birula:CQO7:1996,Birula:PO36:1996}.
However, for $\E,\B \in\RThree$, the wave functions $\G_\pm$ collapse
since in that case $\G_\pm=\G_\mp^\ast$.  This special case has been
studied
\cite{Birula:PO36:1996,Birula&Birula:PRA:2003,Birula&Birula:OC:2006,Berry:JOA:2004a,Berry:JOA:2004b,Kaiser:JOA:2004}
in relation to RS vortices, which are solutions to $(\E+\I{}c\B)^2=0$,
referred to as vortex lines.

\begin{figure}
\includegraphics[width=\columnwidth]{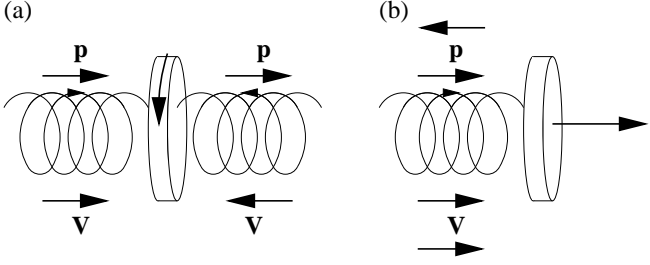}
\caption{%
Right-hand circularly polarized light, depicted by spirals,
impinging from the left onto (a) a $\lambda/2$ plate, (b)
a reflector.  The $\pfield$ arrows indicate the changes in linear
momentum and the $\V=k\s$ arrows indicate
the spin angular momentum changes due to the interactions.
}
\label{fig:plate}
\end{figure}

Let us consider a right-hand circularly polarized wave impinging upon
two different optical elements, as depicted in Fig.~\ref{fig:plate}:
(a) a $\lambda/2$ plate, and (b) a reflector.  According to
Eq.~(\ref{eq:lincon}), the Lorentz force interaction corresponds to a
change in EM linear momentum density, $\Delta\pfield$, and
Eq.~(\ref{eq:spincon}) yields a spin force interaction corresponding to
the change in SAM density, $\Delta\s$.  The interaction with the
$\lambda/2$ plate is an example of a non-Lie transformation; the
corresponding Jones matrix \cite{Allen&al:PRE:1999} has a vanishing
trace.  In this case there is no transfer of linear momentum,
$\Delta\pfield=\0$, but a torque proportional to $\Delta\s=2\s$ is
exerted.  This example corresponds to the Beth experiment
\cite{Beth:PR:1936}.  For the reflector case, $\pfield$, being a polar
vector, changes sign and $\Delta{\pfield}=2\pfield$.  However, $\s$ is a
pseudovector, which does not change sign under reflexion which means
that $\Delta\s=\0$.  In view of this, we propose that Beth's experiments
\cite{Beth:PR:1936} be repeated with elliptically polarized waves.  One
should then observe a torque proportional to $2\s$ rather than
$2\pfield/k$.  If successful, such an experiment would give direct
evidence of the availability of the electromagnetic spin torque density
$\boldsymbol\tau\Sup{spin}$.  The above example could easily be repeated
from an energy conservation perspective based on Eq.~(\ref{eq:Poynting})
and Eq.~(\ref{eq:spin-energy}).

The demonstrated conservation of $H_\pm$ and $\K_\pm$ form the basis of our
theory.  The corresponding conservation laws are quite general, but in
order to give a more complete description of electromagnetic
interactions, we consider classical angular momentum conservation, where
the longitudinal field components play a crucial role.  We take
Eq.~(\ref{eq:GRSmomentum}) as our starting point and cross multiply it
with the position vector $\x$ from the left.  Since
$\tilde{\Tensor{T}}_\pm$ are symmetric, the resulting angular momentum
conservation laws can be written
\begin{align}
\label{eq:POAM}
\pddt{}(\x\cross\K_{\pm})
+\div\left(\x\cross\tilde{\Tensor{T}}_\pm\right)
+\x\cross\fRS_{\pm}=\0\,.
\end{align}
From their sum, one obtains the well-known EM field angular momentum
conservation law
\begin{align}
\label{eq:OAM}
\pddt{\J}+\div{\Tensor{M}}+\x\cross\fLorentz = \0\,,
\end{align}
where $\J=\x\cross\pfield$ is the angular momentum density, and
${\Tensor{M}=\x\times\Tensor{T}}$ is the angular momentum flux tensor
\cite{Schwinger&al:Book:1998}. 
By taking the difference of the two equations in Eq.~(\ref{eq:POAM}),
and introducing $\N=\x\cross\V$, which we interpret as the spin-orbit
angular momentum (SOAM) density, and $\Tensor{O}=\x\cross\Tensor{U}$,
which we interpret as the SOAM flux tensor, we obtain the conservation
law
\begin{align}
\label{eq:SOAM}
\pddt{\N}+\div{\Tensor{O}}+\x\cross\fspin = \0\,.
\end{align}
which, to the best of our knowledge, has not been derived before.
As in solid body mechanics, the spin angular momentum can be viewed as
the intrinsic rotation of the fields and the orbital angular momentum
(OAM) as their precession.  The SOAM would then
correspond to their nutation.  

In analogy with Eq.~(3.30) in Ref.~\onlinecite{Schwinger&al:Book:1998}, one can
derive the virial theorem
\begin{align}
\pddt{(\x\cdot\K_\pm)}+\div(\x\cdot\Tensor{T}_\pm)-H_\pm+\x\cdot\fRS_\pm = 0\,.
\end{align}
Summation and subtraction of the two helicity components yield
\begin{align}
\pddt{(\x\cdot\pfield)}+\div(\x\cdot\Tensor{T})-u+\x\cdot\fLorentz = 0\,,\\
\pddt{(\x\cdot\V)}+\div(\x\cdot\Tensor{U})-w+\x\cdot\fspin = 0\,.
\label{eq:spin-virial}
\end{align}


A commonly accepted interpretation of Beth's experimental results
\cite{Beth:PR:1936} was given in
Ref.~\onlinecite{Simmons&Guttmann:Book:1970} where it is argued that the
finite extent of the waveplate leads to sharp intensity gradients, and
thus strong parallel field components, attributed to a non-vanishing
angular momentum \cite{Allen&Padgett:OC:2000}.  However, a boundary
effect of this kind would be geometry dependent, which is physically
unsatisfactory.  \Eg, in Feynman's example of circularly polarized light
interacting with a free atom \cite{Feynman&al:Book:1965}, it is
difficult to even define a boundary.  Yet, an absorption is followed by
an emission of light with unchanged polarization, just as in our
reflector example in Fig.~\ref{fig:plate}b.  Another example is radio,
where wave polarization can be measured in a single point with an
infinitesimally small antenna.  Hence, SAM can be detected even though
the sensor is much smaller than the wavelength.  In the model presented
here, the result can be explained as a transfer of SAM.  So far, one has
not been able to separate spin angular momentum and orbital angular
momentum other than for beam geometries \cite{Barnett:JOB:2002}.  But
Eqs.~(\ref{eq:spincon}) and (\ref{eq:OAM}) show that, classically,
angular momentum and SAM are conserved independently of each other, so
that the separation is indeed always possible.  An alternative method of
separation is to directly make use of $\G_\pm$, where the spin is
embedded.  The only separation is then with respect to helicity
$\chi=\pm1$.  Thereafter, electromagnetic energy, momentum, and angular
momentum can be unambiguously defined through their respective
conservation laws, Eqs.~(\ref{eq:GRSenergy}), (\ref{eq:GRSmomentum}),
and (\ref{eq:POAM}).

The remaining three Poincar{\'e} invariants are contained in the center
of energy (CE) vector \cite{Boyer:AJP:2005}.  Two CE conservation laws
for positive and negative helicity fields can be derived by multiplying
Eqs.~(\ref{eq:GRSenergy}) with $\x$ and (\ref{eq:GRSmomentum}) with
$c^2t$, which yields 
\begin{multline} \label{eq:RSCE}
\pddt{}\left(\x H_\pm-c^2t\K_\pm\right)
 +c^2\div\left(\x\K_\pm-t\tilde{\Tensor{T}}_\pm\right)  \\
 =\Re{\frac{t\rho}{\epsilon_0}\G_\pm^*\pm\mathrm{i}Z_0 t\j\times\G_\pm^* 
- \mu_0\x(\j\cdot\G_\pm^*)}\,.
\end{multline} 
In vacuum, expressions for the energy and momentum
propagation velocities can be derived \cite{Schwinger&al:Book:1998}.
Assuming them equal, it follows that both right- and left-handed photons
propagate with the speed of light, $c$.  By forming linear combinations
of the CE conservation laws in Eq.~(\ref{eq:RSCE}), the field and spin
CE conservation laws in vacuum are found to be
\begin{gather}
\label{eq:linCE}
\pddt{}\left(\x u-c^2t\pfield\right)
 +c^2\div\left(\x\pfield-t\Tensor{T}\right) = \0\,,
\\
\label{eq:spinCE}
\pddt{}\left(\x v-c^2t\V\right)
 +c^2\div\left(\x\V-t\Tensor{U}\right) = \0\,.
\end{gather}

As demonstrated above, using the framework of generalized RS
electrodynamics we have been able to derive all Poincar{\'e} invariants.
However, there exist other quadratic forms of the RS fields that should be
mentioned.  Reactive but Lorentz invariant observables, obeying
non-conservation laws \cite{Schwinger&al:Book:1998}, can be derived by
examining forms which are quadratic in $\G_\pm$ and $\G_\mp^\ast$.  The
Lorentz scalars $\epsz(\E\cdot\E^{\ast}-c^2\B\cdot\B^{\ast})/2$ and
$\Re{\E\cdot\B^\ast}/Z_0$, and the imaginary part of the complex linear
momentum vector $\epsz\Im{\E\cross\B^\ast}$ are important examples.
Similarly, ``instantaneous'' quantities, conserved and non-conserved,
can be derived by considering forms quadratic in $\G_\pm$, $\G_\mp$, and
$\G_\pm$, $\G_\pm$, respectively.  The theory can be generalized to
incorporate a magnetic charge density, $\rho\Sub{m}$, and current
density, $\j\Sub{m}$, by introducing $\rho_\pm=\rho\Sub{e}\pm
\I\rho\Sub{m}/c$ and $\j_\pm=\j\Sub{e}\pm \I\j\Sub{m}/c\,$
\cite{Ibragimov&al:JMP:2007}, sometimes referred to as the Beltrami
charge and current densities \cite{Weiglhofer&Lakhakia:PRE:1994}.

In conclusion, based on the assumption of an analytic continuation of
the fields so that $\E,\B\in\CThree$, we have introduced generalized RS
vectors $\G_\pm$ that are interpreted as wave functions describing
fields of positive and negative helicity.  This has allowed us to
derive the two sets of Poincar{\'e} invariants
$\left\{H_\pm,\K_\pm,\x\cross\K_\pm,(\x H_\pm-c^2t\K_\pm)\right\}$ and
their associated conservation laws, Eqs.~(\ref{eq:GRSenergy}),
(\ref{eq:GRSmomentum}), (\ref{eq:POAM}), and (\ref{eq:RSCE}),
respectively.  All well-known EM observables are contained in these sets
as linear combinations of the two versions, but some are less well known
or seem to have gone unnoticed.  Eq.~(\ref{eq:spin-energy}), the
spin-energy equivalent to the Poynting theorem Eq.~(\ref{eq:Poynting}),
are among these.  The SOAM conservation law, Eq.~(\ref{eq:SOAM}), and
the spin CE conservation law Eq.~(\ref{eq:spinCE}), as well as the spin virial
theorem Eq.~(\ref{eq:spin-virial}), are, to the best of
our knowledge, given here for the first time.

The authors thank Nail Ibragimov, Gunnar Ingelman, and Oscar St{\aa}l
for valuable comments.  This work was financially supported by the
Swedish National Space Board and the Swedish Research Council.

\bibliographystyle{ol}
\bibliography{oam}

\end{document}